\documentclass[prl,twocolumn,amsmath,amssymb]{revtex4}
\usepackage{graphicx}
\usepackage{dcolumn}
\usepackage{bm}

\newcommand{\om}{\omega}
\newcommand{\omp}{\omega^{\prime}}
\newcommand{\omc}{\omega_{c}}
\newcommand{\omps}{\omega^{\prime 2}}
\newcommand{\sr}{\sigma_{1}}
\newcommand{\si}{\sigma_{2}}
\newcommand{\omin}{\omega_{\mbox{\scriptsize{min}}}}
\newcommand{\omax}{\omega_{\mbox{\scriptsize{max}}}}

\begin{document}



\title{Model-Independent Sum Rule Analysis Based on Limited-Range Spectral Data}

\author{A.B. Kuzmenko}
\author{D. van der Marel}
\author{F. Carbone}
\author{F. Marsiglio} \altaffiliation[Permanent affiliation:
]{ Dept. of Physics, University of Alberta, Edmonton, AB,
Canada, T6G 2J1.}

\affiliation{DPMC, University of Geneva, 1211 Geneva 4,
Switzerland\\}

\begin{abstract}

Partial sum rules are widely used in physics to separate low-
and high-energy degrees of freedom of complex dynamical
systems. Their application, though, is challenged in practice
by the always finite spectrometer bandwidth and is often
performed using risky model-dependent extrapolations. We show
that, given spectra of the real and imaginary parts of any
causal frequency-dependent response function (for example,
optical conductivity, magnetic susceptibility, acoustical
impedance etc.) in a limited range, the sum-rule integral from
zero to a certain cutoff frequency inside this range can be
safely derived using only the Kramers-Kronig dispersion
relations without any extra model assumptions. This implies
that experimental techniques providing both active and reactive
response components independently, such as spectroscopic
ellipsometry in optics, allow an extrapolation-independent
determination of spectral weight 'hidden' below the lowest
accessible frequency.

\pacs{ PACS numbers: 78.20.Bh,74.25.Gz,02.30.Zz,02.60.-x}
\end{abstract}

\maketitle

\section{Introduction}

Global sum rules applied to response functions play a major
role in physics as they quantitatively express fundamental
conservation laws. Of interest are also {\em partial}, or {\em
restricted} sum rules, where a properly chosen cutoff frequency
separates low- and high-energy degrees of freedom of a physical
system. For example, the low-frequency optical spectral weight
\cite{Wooten}
\begin{equation}
W(\omc)=\int_{0}^{\omc}\sigma_{1}(\om)d\om\label{SW}
\end{equation}
\noindent is a partial counterpart of the well known $f$-sum
rule $\int_{0}^{\infty}\sigma_{1}(\om)d\om=\pi n e^2/(2m_{e})$
for the optical conductivity $\sigma(\omega)=\sigma_{1}(\omega)
+ i \sigma_{2}(\omega)$, where $n$ is the density of charges,
$e$ and $m_{e}$ are the charge and the bare mass of electron.
In charge conducting systems, the integral up to a cut-off
frequency $\omega_{c}$ somewhat larger than the free-carrier
scattering rate but below the energies of transitions from
occupied to empty bands is proportional to the number of
carriers and the inverse band mass averaged over the Fermi
surface. For example, in the simple Drude model, integrating
out to 10 (20) times the scattering rate recovers 94\% (97\%) of the sum
rule.

The effective mass can be strongly affected by
electron correlations, especially in a case of narrow
bandwidth. The dependence of $W(\omega_{c})$ on
temperature and other parameters, especially across a phase
transition, is thus a valuable piece of information about the
changes in the electronic system. The changes of
$W(\omega_{c})$ can be very large, as at the ferromagnetic -
paramagnetic transition in colossal magneto-resistance
manganites \cite{OkimotoPRL95,MillisJESP01}, or rather small as
at the superconducting transition in the high-$T_c$ cuprates
\cite{BasovScience99,MolegraafScience02,SantanderEPL03,BorisScience04}.
Even in the latter case, the subtle variations of the
low-frequency spectral weight may potentially distinguish
between physically different scenarios of superconductivity
\cite{HirschPC92}. If a significant spectral overlap between the free
charge and interband peaks is present, the intrepretation can
be less obvious and requires direct theoretical calculation of
the value of $W(\omc)$. Therefore it is worth using all means,
experimental and computational, to improve the accuracy of
$W(\omega_{c})$ determined from the available spectra. Notably,
the partial sum rules can be meaningfully applied not
only to the optical spectra, but also in acoustical data,
neutron scattering and other spectroscopic techniques.

Because of limitations on the bandwidth of any spectrometer,
$\sigma_{1}(\omega)$ is not experimentally available down to
zero frequency. The direct application of Eq.(\ref{SW})
assuming some low-frequency extrapolations of
$\sigma_{1}(\omega)$ may lead to significant and uncontrollable
error bars. This is most obvious for the optical conductivity
of a superconductor, where the spectral weight of the
condensate of the Cooper pairs is represented in
$\sigma_{1}(\omega)$ by a $\delta$-peak at zero-frequency. Here
the inductive component $\sigma_{2}(\omega)$ has to be used to
estimate the condensate spectral weight
\cite{TinkhamPR56,DordevicPRB02}. However, the common procedure
used in this case still requires the extrapolation down to zero
frequency.

Importantly, certain experimental techniques, such as spectroscopic ellipsometry,
or simultaneous measurement of acoustical attenuation and the sound speed allow direct
independent measurement of both components of the response function. The purpose of this article is to
show that the spectral weight $W(\omega_{c})$, including a possible zero-frequency
$\delta$-peak, can be obtained model-independently, {\em i.e.}
without any {\em a priori} assumptions about the low- and
high-frequency spectral behavior, if both $\sigma_{1}(\omega)$
and $\sigma_{2}(\omega)$ are measured in a {\em limited}
frequency range [$\omin$, $\omax$]. We also present an efficient
numerical algorithm optimized to reduce the output error bars in
the case of the noisy data.

\section{On the possibility of analytical continuation from a finite interval}

The causality principle, {\em i.e.} the assumption that no
response can precede the causing factor, puts constraints on
the analytic behavior of all physical response functions, such
as the optical conductivity $\sigma(\omega)=\sigma_{1}(\omega)
+ i \sigma_{2}(\omega)$. In particular, these functions obey
the Kramers-Kronig (KK) dispersion relations:
\begin{eqnarray}
\sr(\om)&=&\frac{2}{\pi}\wp\int_{0}^{\infty}\frac{\omp\si(\omp)d\omp}{\omps-\om^2}\label{KK1}\\
\si(\om)&=&\frac{2\om}{\pi}\wp\int_{0}^{\infty}\frac{\sr(\omp)d\omp}{\om^2-\omps}\label{KK2}
\end{eqnarray}
\noindent where $\wp$ denotes the principal-value integral.
Note that relations (\ref{KK1}) and (\ref{KK2}) are entirely
model-independent since they only use the fact that the
response function is analytical in the upper complex semiplane.
One can say that the information about $\sigma_{1}(\omega)$ is
encoded in $\sigma_{2}(\omega)$ and {\it vice versa}.

The determination of the partial sum-rule integral (\ref{SW})
is intimately related to a more general problem to restore the
function $\sigma(\omega)$ itself outside the experimental
range. It is well known \cite{ComplexAnalysis} that a complex
function $\sigma(\om)$ analytical (holomorphic) in a certain
domain $D$ can be analytically continued from a subset $\Gamma$
of the boundary of this domain into the whole domain, including
the rest of the boundary. The specific form of such a
continuation has been a subject of numerous studies since the
late 1920's \cite{Carleman26}, which are summarized in
Ref.\onlinecite{AizenbergKluwer93}. In particular, the
Carleman-Goluzin-Krylov formulas restore the function {\em
exactly}\cite{RemarkCarleman} inside the analyticity domain from its values at
$\Gamma$. They have the following general structure:
\begin{equation}\label{CGK}
\sigma(\om)=\lim_{n\rightarrow\infty}\int_{\Gamma}
Q_{n}(\omp,\om)\sigma(\omp)d\omp, \ \ \ \om\in D.
\end{equation}
\noindent There are several possibilities to choose the kernel
$Q_{n}(\omp,\om)$. We point out just one option for the case
$\Gamma = [\omin,\omax]$ \cite{AizenbergKluwer93}
\begin{equation}\label{KernelQ}
Q_{n}(\omp,\om)= \frac{(2\pi i)^{-1}}{\omp -
\om}\left[\frac{(\omp-\omin)(\om-\omax)}{(\omp-\omax)(\om-\omin)}\right]^{\frac{n
i }{\pi}}
\end{equation}

\noindent that illustrates the general property of these
kernels: they oscillate as a function of both variables, with
the frequency of oscillations increasing infinitely as $n$
increases. Therefore the 'lim' operation cannot be
applied directly to $Q_{n}(\omp,\om)$ as it only exists for the
whole integral (\ref{CGK}). Using Eqs. (\ref{CGK}) and (\ref{KernelQ}) one can, in principle, obtain the spectral
weight
\begin{equation}\label{CGKSW}
W(\omc)=\mbox{Re
}\left\{\lim_{n\rightarrow\infty}\int_{\omin}^{\omax}
u_{n}(\omp,\omc)\sigma(\omp)d\omp\right\}
\end{equation}
\noindent where
$u_{n}(\omp,\omc)=\int_{0}^{\omc}Q_{n}(\omp,\om)d\om$.

\begin{figure}[htb]
   \centerline{\includegraphics[width=9cm,clip=true]{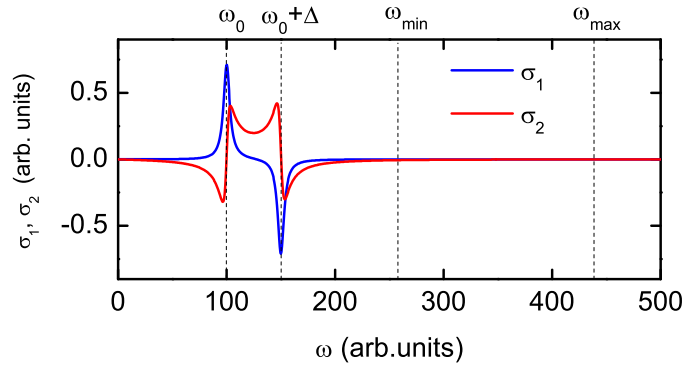}}
   \caption{(Color online) An example of $\sigma_{1}(\omega)$ and $\sigma_{2}(\omega)$
   satisfying KK relations which are both almost zero
   at $[\omin,\omax]$ but show intense spectral features below $\omin$. The 'unphysical' negative values of $\sigma_{1}(\omega)$ can be meaningful,
   for example, if $\sigma_{1}(\om)$ is an addition to some background response function so that the result is positive. }
   \label{Fig0}
\end{figure}

In spite of the fact of its formal existence, the analytical
continuation of a function from a finite interval appears to be
a very ill-posed problem. For example, one can construct an
analytical function, which is almost zero in the range
$[\omin,\omax]$ but shows intense spectral structure below
$\omin$ (Fig. \ref{Fig0}). Let us consider $\sigma(\omega)$ to
be a sum of two narrow Lorentzians: the first one with spectral
weight $A$ centered at $\om_{0}$ below $\omin$ and the second
one with the opposite spectral weight $-A$ slightly displaced
to $\om_{0}+\Delta$, still below $\omin$. Normally,
$\sigma_1(\om)$ cannot be negative; however, in this example
one can think of $\sigma_1(\om)$ as an addition to some
positive background response function, which makes the result
always positive. A similar function would appear, in particular, if one takes a
difference between two optical conductivities of the same sample at two different
temperatures in a case when the conductivity contains a single optical
phonon peak which shifts as a function of temperature (assuming that the
phonon spectral weight remains unchanged).The width of the peaks is assumed to be much
less than $\Delta$). The corresponding $\sigma_{2}(\omega)$ far
from $\om_{0}$ would be approximately
$(-4A\om_{0}\om\Delta/\pi)(\om^2-\om_{0}^2)^{-2}$. By
decreasing $\Delta$, both $\sigma_{1}(\omega)$ and
$\sigma_{2}(\omega)$ can be made vanishingly small in the range
$[\omin,\omax]$. Obviously, the slightest noise on top of
$\sigma(\omega)$ that would make it indistinguishable from zero
prevents the extraction of the mentioned strong structures
beyond the accessible range.

In the considered example it was essential that the spectral
weights of the two peaks exactly compensate each other in order
to get the vanishing values of $\sigma_{1}(\omega)$ and
$\sigma_{2}(\omega)$ at $[\omin,\omax]$. Otherwise one would
get a detectable term $\sim 1/\om$ in $\sigma_{2}(\om)$
proportional to the total low-frequency spectral weight
\cite{BozovicPRB90,KuzmenkoPRB05}. This indicates that the
'hidden' spectral weight is much better determined by the
limited-range data than the function $\sigma_{1}(\omega)$
itself. This is supported by the study of Aspnes
\cite{AspnesPRB75}, who examined the possibility to extrapolate
an ellipsometrically measured dielectric function beyond the
experimental range using the Kramers-Kronig relations and found
that the total spectral weights of a few broad spectral regions
can be restored reasonably well while the high-resolution
details of the spectra cannot be unambiguously determined.
Some analytical treatments of the problem of the finite frequency
range can be found in Refs. \cite{HulthenJOSA72,MiltonPRL97,DienstfreyIP01}.

\section{A practical algorithm for the noisy data}

We are interested in an efficient and accurate numerical scheme to
determine the sum-rule integral (\ref{SW})). It turns out that the
straightforward application of the formulas (\ref{CGK}) and
(\ref{CGKSW}) with strongly oscillating kernels to real data is
not practical as it amplifies uncontrollably the experimental
noise. In this case one has to look for different numerical
algorithms.

The experimental spectra are collections of data points
$\sigma_{1,j}\pm\delta\sigma_{1,j}$ and
$\sigma_{2,j}\pm\delta\sigma_{2,j}$ on a mesh of frequencies
$\om_{j}$ ($j=1,..,N$). We assume that the spectral resolution
is roughly the same as the distance between neighboring points.
According to Eq. (\ref{CGKSW}), $W(\omc)$ is a linear function
of the real and imaginary parts of $\sigma$. We note that this
directly follows from the fact that the KK relations are linear
and is independent from the particular scheme of analytical
continuation. Hence it is logical to take the following formula
for the calculations:
\begin{equation}\label{Wlinsum1}
W(\omc)\approx\sum_{j=1}^{N}\left[u_{1,j}\sigma_{1,j}+u_{2,j}\sigma_{2,j}\right],
\end{equation}
\noindent which is the most general linear relation between
$W(\omc)$ and the measured values $\sigma_{1,j}$ and
$\sigma_{2,j}$. The coefficients $u_{1,j}$ and $u_{2,j}$ that
we call hereafter u-coefficients have to be chosen in such a
way that formula (\ref{Wlinsum1}) is approximately correct for
any {\em arbitrarily} chosen response function. Since any
causal response function can be represented as a linear
superposition of narrow oscillator response functions
$S_{x}(\om)=S_{1x}(\om)+i S_{2x}(\om)$ centered at all
frequencies $x$, one should optimize the u-coefficients in such
a way that it gives a reasonably accurate answer when applied
to $\sigma(\om)=S_{x}(\om)$ for {\em any} $x$. For
$S_{1x}(\om)$, one can take a narrow peaked function, for
example a Gaussian, while $S_{2x}(\om)$ should be the KK
transform of $S_{1x}(\om)$.

We introduce a function $D(x)$, which is the inaccuracy of the
formula (\ref{Wlinsum1}) when applied to $S_{x}(\om)$
\begin{eqnarray}
D(x)=\int_{0}^{\omc}S_{1x}(\om)d\om-
\sum_{j=1}^{N}\left[u_{1,j}S_{1x}(\om_{j})+u_{2,j}S_{2x}(\om_{j})\right]\nonumber
\end{eqnarray}
\noindent The integrated inaccuracy can be defined as follows
\begin{eqnarray}
D_{int}^{2}=\int_{0}^{\infty}D^{2}(x)dx.
\end{eqnarray}
\noindent In order to make the formula (\ref{Wlinsum1})
accurate for all $x$ at the same time, one would need to
minimize $D_{int}^{2}$ by varying the u-coefficients. If
experimental noise is present, one should also take care that
the error bar of the resulting spectral weight due to the noise
\begin{eqnarray}
F^{2}=\sum_{j=1}^{N}\left[u_{1,j}^2(\delta\sigma_{1,j})^2+u_{2,j}^2(\delta\sigma_{2,j})^2\right]
\end{eqnarray}
\noindent does not become too large. In general, $D_{int}^{2}$
and $F^{2}$ have to be minimized simultaneously. A more detailed description of this algorithm is given in the Appendix 1.

\begin{figure}[htb]
   \centerline{\includegraphics[width=6.8cm,clip=true]{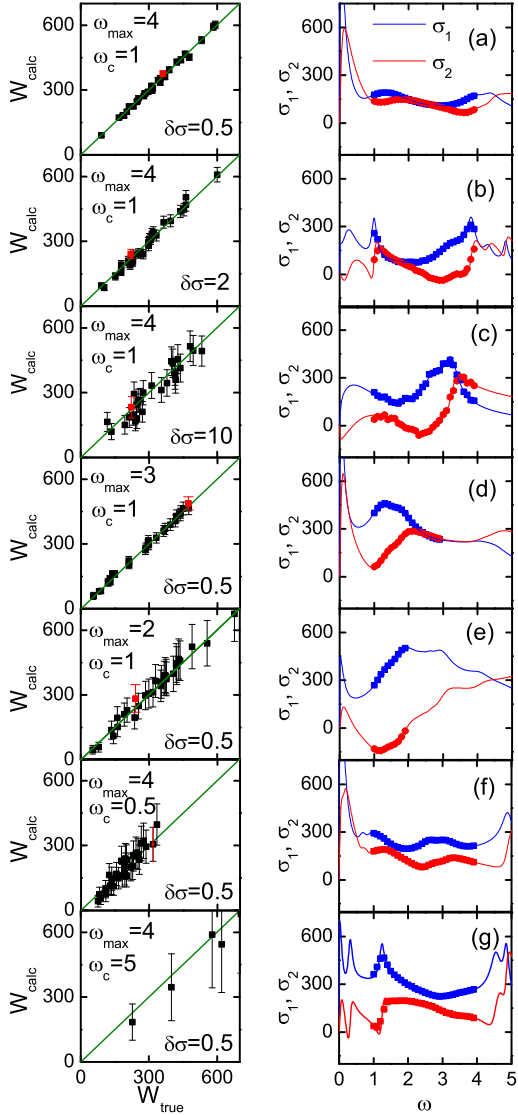}}
   \caption{(Color online) Series (a)-(g) of Monte-Carlo tests of the numerical algorithm that calculates the spectral weight $W(\omc)$
   knowing only $\sigma_{1}(\omega)$ and $\sigma_{2}(\omega)$ in a limited range $[\omin, \omax]$.
   Left panel: the correlation between the 'true' spectral weight $W_{true}(\omc)$ and the
   calculated one $W_{calc}(\omc)$; each point corresponds to one
   randomly generated KK consistent response function as described in the
   text. The error bars are as estimated by the program. Green lines refer to
   $W_{calc}(\omc)=W_{true}(\omc)$. Right panel: the curves $\sigma_{1,2}(\om)$ corresponding
   to the point marked in red color on the left panel. Solid
   lines are the 'true' functions, dots are the data points (with noise $\delta\sigma$ added and finite resolution applied) used as input.
   The values of $\omax$, $\omc$ and $\delta\sigma$ corresponding to each series are given on the left panels ($\omin=1$ in all series).}
   \label{fig1}
\end{figure}

We implemented this idea in a numerical code Devin
\cite{Devin}, which takes a set of data points of
$\sigma_{1}(\om)$ and $\sigma_{2}(\om)$ with error bars in a
limited range $[\omin,\omax]$ and returns the estimated value
and error bar of $W(\omc)$ for specified cutoff frequency
$\omc$. It is assumed that the real and imaginary parts of the
response function are KK-consistent; otherwise output error bars are
unpredictable. Note that the KK consistency of the
limited-range data can be tested using exact bounds proposed in
Ref.\cite{MiltonPRL97}.

\section{Random tests of the method}

A series of Monte-Carlo tests were performed where the program
answer based on limited-range spectral information can be
compared with the exact value of $W(\omc)$, allowing one to
scrutinize the model independence of the method. We generate a
random KK-consistent function $\sigma(\omega)$ by adding up a
random number of peaks of random spectral weights $A_{k}$,
widths $\gamma_{k}$ and center frequencies $\om_{0,k}$
distributed below, inside and above the range $[\omin,\omax]$.
In particular, a sum of Lorentz peaks plus high-frequency
background was taken\cite{units}: $\sigma(\om)= \sum_{k}(-2i
A_{k}\om/\pi)(\om_{0,k}^2-\om^2-i\gamma_{k}\om)^{-1} -
i\epsilon_{\infty}\om/(4\pi)$, where the number of peaks is
changed between 1 and 50. The parameters were varied in the
following limits: $A_{k}$ - between 0 and 125, $\gamma_{k}$ -
between 0 and 3, $\omega_{0,k}$ - between 0 and 5,
$\epsilon_{\infty}$ - between 0 and 5. The first oscillator was
always a delta function at zero frequency ($\om_{0,1}=0$,
$\gamma_{1}=0$), imitating a condensate peak in
superconductors. The 'experimental' data points sent to the
program were generated with a step of 0.1 inside the range
$[\omin,\omax]$ by convoluting the true functions with a
Gaussian of width 0.1 to mimic the finite resolution and adding
random noise with standard deviation of $\delta\sigma$. Each
test series consisted of 30 independent generations of
$\sigma(\omega)$ and a comparison of the true and calculated
$W(\omc)$.

The test results are summarized in Fig.\ref{fig1}. The series
(a)-(c) demonstrate the sensitivity of the extracted spectral
weight to the data noise. The experimental range [1, 4] and the
cutoff $\omc=1$ are the same, while $\delta\sigma$ varies from 0.5
to 10. One can see that for sufficiently small noise the
program is able to provide quite accurate value of $W(\omc)$
for all 30 random inputs, which demonstrates that the method is
indeed model-independent. On the other hand, relatively large
error bars in series (b) and (c) show that the requirements to
the signal-to-noise ratio for this particular set of data
points are quite strict ($\sim$ 1 \%), although not
unrealistic.

Series (a), (d) and (e) give a feeling of how the accuracy of
this procedure depends on the width of experimental range at a
constant noise amplitude. The range was consecutively narrowed
from [1,4] to [1,2]. The range width appears to be a critical
factor determining the method precision. The
output error bars increase rapidly as we restrict the
experimental range - much faster, in fact, than what one would
normally expect due to a simple decrease of the number of data
points.

By comparing series (a), (f) and (g) one can see that the error
bars increase dramatically if the cutoff $\omc$ is taken beyond
the experimental range, both below $\omin$ or above $\omax$.
This is another indication that one can reliably determine the
total 'hidden' spectral weight but not a part of it since the
latter depends on the inaccessible spectral details of
$\sigma_{1}(\om)$.

\begin{figure}[htb]
   \centerline{\includegraphics[width=7cm,clip=true]{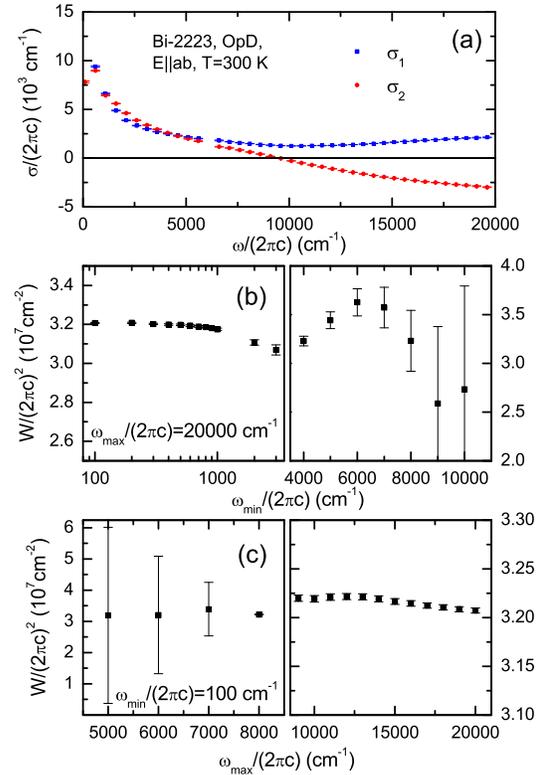}}
   \caption{(Color online) The application of the partial sum-rule analysis to optical
   conductivity (a) of optimally doped Bi$_{2}$Sr$_{2}$Ca$_{2}$Cu$_{3}$O$_{10}$ at
   room temperature for electric field parallel to CuO$_{2}$ planes (Ref.\onlinecite{CarbonePRB06}).
   All frequencies and conductivities are divided by $2\pi c$ in order to get the wavenumber
   units of cm$^{-1}$. Above 6000 cm$^{-1}$, $\sigma_{1}(\om)$ and $\sigma_{2}(\om)$ were measured directly by
   spectroscopic ellipsometry; below 6000 cm$^{-1}$ they were
   derived with error bars from simultaneous KK-constrained fit of reflectivity \cite{KuzmenkoRSI05}, measured
   down to 100 cm$^{-1}$ and the ellipsometric spectra. (b) The value of $W(\omc)$ calculated for $\omc$=8000
   cm$^{-1}$ as a function of $\omin$ for $\omax$=20000 cm$^{-1}$. (c) The value of $W(\omc)$ for $\omc$=8000 cm$^{-1}$
   as a function of $\omax$ for $\omin$=100 cm$^{-1}$. The error bars are indicative, as estimated by the program.}
   \label{FigExample}
\end{figure}

\section{Real-data example}

In Fig.\ref{FigExample} the application of this program to real
data is demonstrated. As an example, we calculate the sum-rule
integral for the optical conductivity of a high-$T_{c}$
superconducting compound
Bi$_{2}$Sr$_{2}$Ca$_{2}$Cu$_{3}$O$_{10}$, based on data
published in Ref.\onlinecite{CarbonePRB06}. We use a set of
data which spans the interval of wavenumbers $\om/(2\pi c)$
between 100 cm$^{-1}$ and 20000 cm$^{-1}$. In
Fig.\ref{FigExample}b each point corresponds to a value of
$W(\omc)$ calculated on the basis of input data from $\omin$ to
$\omax$=20000 cm$^{-1}$ as a function of $\omin$. The error
bars are relatively small as long as $\omin$ is less than
$\omc$, but grow explosively as $\omin$ exceeds $\omc$. In
Fig.\ref{FigExample}c each point corresponds to a value of
$W(\omc)$ calculated on the base of input data from $\omin$=100
cm$^{-1}$ to $\omax$ as a function of $\omax$. When $\omc$ is
above the highest frequency point, the error bars again tend to
diverge.

\section{Discussion and conclusions}

One should note that the the involved algorithm that we used to
extract the partial sum-rule integral $W(\omc)$ is not the only possibility
to tackle this
problem. Perhaps a more conventional and intuitive method is to
fit the available spectra with a KK-consistent multi-parameter
function, such as the Drude-Lorentz model or the variational
function of Ref.\cite{KuzmenkoRSI05}. A similar method of the
Kramers-Kronig integrals was used by Aspnes \cite{AspnesPRB75}.

Interestingly, all fitting schemes where the model function is
a linear superposition of the trial basis functions eventually
reduce to the same type of linear formula (\ref{Wlinsum1}), as
demonstrated in the Appendix 2. However, there is an important
difference between the two approaches. In the method that we
used in this paper, the u-coefficients are optimized in order
to minimize the output error bars $\delta W(\omc)$, while in
the fitting approach they are predetermined by the specific set
of trial basis functions, which may not be always optimal. In
this sense, our approach is more general and model-independent.

On another hand, the data fitting approach can be
straightforwardly applied also to experimental quantities
that depend on $\sigma_{1}(\om)$ and
$\sigma_{2}(\om)$ in a non-linear way, for example, the optical reflectivity.
Another advantage of the data fitting technique is
that it can detect if $\sigma_{1}(\om)$ and $\sigma_{2}(\om)$
are KK inconsistent, for example, due to the systematic
experimental uncertainties. Thus it is preferable to use both techniques
in a combination.

In summary, we have shown that the partial sum-rule analysis
can be accurately performed on the basis of the real and imaginary
parts of a response function, if the latter is experimentally
available in a limited spectral range. In a sense, nature
integrates for us $\sigma_{1}(\om)$ beyond the accessible range
and encodes information about this integral in
$\sigma_{2}(\om)$ inside the range where the experimental data
is available. We have shown that it can be decoded using a
simple linear formula (\ref{Wlinsum1}) with optimally chosen
u-coefficients. The determination is accurate only if the
cutoff frequency is lying inside the accessible interval. Even
though the interval can be, formally speaking, arbitrarily
small, the extraction of the sum-rule integral for a narrow
interval would require much better data accuracy to obtain
equivalent precision of $W(\omc)$ than for a broad one. We
find, however, that the error bars are not uncontrollably large
due to the notorious extrapolation uncertainty. They can be, at
least in principle, made arbitrarily small by decreasing the
experimental noise. This result is valid for all response
functions satisfying the KK relations and thus applies to
various domains of spectroscopy.

\section{Acknowledgements}

This work was supported by the Swiss National Science
Foundation through the National Center of Competence in
Research "Materials with Novel Electronic Properties-MaNEP".
The work by FM was supported by NSERC, CIAR and ICORE. We
acknowledge fruitful discussions with H. Molegraaf, E. van
Heumen, C. Bernhard, D. Basov, R.K. Teshima, S.V. Rotin, A.B.
Bogatyrev and G. Savar\`e.

\section{Appendix 1. Optimization of the u-coefficients}\label{Section1}

\

Here we describe a method that we use to optimize the
u-coefficients in the approximate formula
\begin{equation}\label{Wlinsum}W(\omc)\approx W_{a}=\sum_{\nu=1}^{2}\sum_{j=1}^{N}u_{\nu,j}\sigma_{\nu,j}.
\end{equation}

\noindent The starting point is the spectral representation of
the response function:
\begin{equation}\label{SpecRep}
\sigma(\om)= \wp\int_{0}^{\infty}\sr(x)S(\om,x)dx,
\end{equation}
\noindent where $S(\om,x)=\delta(\omega+x)+\delta(\omega-x)+
(i/\pi)[(\om-x)^{-1}+(\om+x)^{-1}]$ is a response of an
oscillator centered at frequency $x$ (and at $-x$ to preserve
the parity of the response functions). Put differently, any
causal response function can be represented as a linear
superposition of narrow oscillator functions. In reality, one
has to convolute $S(\om,x)$ with an apparatus function
$A(\om,\om')$, for example, a Gaussian, which takes the
experimental resolution into account: $\tilde{S}(\om,x)=\int
A(\om,\om') S(\om',x)d\om'$. The key idea of the method is
based on the linearity of Eq. (\ref{Wlinsum}): in order to make
this formula applicable to any $\sigma(\om)$, one should
optimize the u-coefficients in such a way that it gives a
reasonably accurate answer when applied to
$\sigma(\om)=\tilde{S}(\om,x)$ for {\em any} $x$.

To be quantitative, one can calculate a discrepancy function
$D(x)$ for a given combination of the u-coefficients
\begin{eqnarray}
D(x)=\tilde{W}_{s}(x)-\tilde{W}_{s,a}(x)\nonumber,
\end{eqnarray}

\noindent where
\begin{eqnarray}
\tilde{W}_{s}(x)&=&\int_{0}^{\omc}\tilde{S}_{1}(\om,x)d\om\nonumber\\
\tilde{W}_{s,a}(x)&=&\sum_{\nu=1}^{2}\sum_{j=1}^{N}u_{\nu,j}\tilde{S}_{\nu}(\om_{j},x)\nonumber
\end{eqnarray}
\noindent One can also define the discrepancy integrated over
all $x$
\begin{equation}\label{F}
D_{int}^{2}=\int_{0}^{\infty}D^{2}(x)dx.
\end{equation}
\noindent In order to make the formula (\ref{Wlinsum}) accurate
for all $x$ simultaneously (assuming that it is possible!), one
should minimize the integral discrepancy by varying the
u-coefficients. This is easy since
$D_{int}^{2}(\{u_{1,j}\},\{u_{2,j}\})$ is a quadratic form
\begin{eqnarray}
D_{int}^{2}=\sum_{\nu,\mu=1}^{2}\sum_{i,j=1}^{N}A_{\nu\mu,ij}u_{\nu,i}u_{\mu,j}
           -2\sum_{\nu=1}^{2}\sum_{i=1}^{N}B_{\nu,i}u_{\nu,i}+C\nonumber
\end{eqnarray}
\noindent where the coefficients are given by the integrals
over $x$:
\begin{eqnarray}\label{coefs}
A_{\nu\mu,ij}&=&\int_{0}^{\infty}\tilde{S}_{\nu}(\om_{i},x)\tilde{S}_{\mu}(\om_{j},x)dx,\nonumber\\
B_{\nu,i}&=&\int_{0}^{\infty}\tilde{S}_{\nu}(\om_{i},x)\tilde{W}_{s}(x)dx,\nonumber\\
C&=&\int_{0}^{\infty}\tilde{W}_{s}^{2}(x)dx.
\end{eqnarray}

\noindent The minimization of $D_{int}^{2}$ reduces to the
linear system of equations:
\begin{eqnarray}
\sum_{\mu=1}^{2}\sum_{j=1}^{N}A_{\nu\mu,ij}u_{\mu,j}=B_{\nu,i}\nonumber
\end{eqnarray}

In a case when the (frequency-dependent) spectral resolution
$\delta(\om)$ is given by a Gaussian
\begin{eqnarray}
A(\om,\om')=\frac{1}{\sqrt{2\pi}\delta(\om)}\exp\left[-\frac{(\om-\om')^{2}}{2\delta^{2}(\om)}\right]\nonumber
\end{eqnarray}
\noindent the above integrals can be taken analytically, which
simplifies dramatically the calculations:
\begin{eqnarray}
\tilde{S_{1}}(\om,x)&=&\frac{1}{\sqrt{2}\delta(\om)}\left[
g_{1}\left(\frac{\om+x}{\sqrt{2}\delta(\om)}\right)+
g_{1}\left(\frac{\om-x}{\sqrt{2}\delta(\om)}\right)\right]\nonumber \\
\tilde{S_{2}}(\om,x)&=&\frac{1}{\sqrt{2}\delta(\om)}\left[
g_{2}\left(\frac{\om+x}{\sqrt{2}\delta(\om)}\right)+
g_{2}\left(\frac{\om-x}{\sqrt{2}\delta(\om)}\right)\right]\nonumber \\
\tilde{W}_{s}(x)&=&\frac{1}{2}\left[
g_{3}\left(\frac{\omc+x}{\sqrt{2}\delta(\om_{c})}\right)+
g_{3}\left(\frac{\omc-x}{\sqrt{2}\delta(\om_{c})}\right)\right]\nonumber
\end{eqnarray}
\noindent and
\begin{eqnarray}
A_{11,ij}&=&\frac{1}{\delta_{ij}}\left[
g_{1}\left(\frac{\om_{ij}^{+}}{\delta_{ij}}\right)+g_{1}\left(\frac{\om_{ij}^{-}}{\delta_{ij}}\right)\right]\nonumber\\
A_{12,ij}&=&\frac{1}{\delta_{ij}}\left[
g_{2}\left(\frac{\om_{ij}^{+}}{\delta_{ij}}\right)-g_{2}\left(\frac{\om_{ij}^{-}}{\delta_{ij}}\right)\right]\nonumber\\
A_{21,ij}&=&\frac{1}{\delta_{ij}}\left[
g_{2}\left(\frac{\om_{ij}^{+}}{\delta_{ij}}\right)+g_{2}\left(\frac{\om_{ij}^{-}}{\delta_{ij}}\right)\right]\nonumber\\
A_{22,ij}&=&\frac{1}{\delta_{ij}}\left[
-g_{1}\left(\frac{\om_{ij}^{+}}{\delta_{ij}}\right)+g_{1}\left(\frac{\om_{ij}^{-}}{\delta_{ij}}\right)\right]\nonumber
\end{eqnarray}
\begin{eqnarray}
B_{1,i}&=&g_{3}\left(\frac{\om_{ic}^{+}}{\delta_{ic}}\right)-g_{3}\left(\frac{\om_{ic}^{-}}{\delta_{ic}}\right)\nonumber\\
B_{2,i}&=&g_{4}\left(\frac{\om_{ic}^{+}}{\delta_{ic}}\right)-g_{4}\left(\frac{\om_{ic}^{-}}{\delta_{ic}}\right)\nonumber\\
C&=&\delta_{c}g_{5}\left(\frac{\om_{c}}{\delta_{c}}\right)\nonumber
\end{eqnarray}

\noindent where $\om_{ij}^{\pm}=(\om_{i}\pm\om_{j})/2$,
$\delta_{ij}^{2}=[\delta^{2}(\om_{i})+\delta^{2}(\om_{j})]/2$,
$\om_{ic}^{\pm}=(\om_{i}\pm\om_{c})/2$,
$\delta_{ic}^{2}=[\delta^{2}(\om_{i})+\delta^{2}(\om_{c})]/2$
and $\delta_{c}=\delta(\om_{c})$. Here we used auxiliary
functions
\begin{eqnarray}
g_{1}(x) &=& \pi^{-1/2}\exp(-x^{2})\nonumber\\
g_{2}(x) &=& \pi^{-1/2}\exp(-x^{2})\mbox{ erfi}(x)\nonumber\\
g_{3}(x) &=& \mbox{ erf}(x)\nonumber\\
g_{4}(x) &=& 2\pi^{-1}x^{2}\ {}_{2}F_{2}(1,1,3/2,1,-x^2)\nonumber\\
g_{5}(x) &=& 2\pi^{-1/2}[\exp(-x^{2})-1]+2x\mbox{
erf}(x)\nonumber,
\end{eqnarray}

\noindent where $\mbox{ erf}(x)$ is the error function, $\mbox{
erfi}(x)=\mbox{erf}(ix)/i$ the imaginary error function and
${}_{2}F_{2}(..,x)$ the hypergeometric function.

So far we ignored the experimental uncertainty of
$\sigma_{1}(\om)$ and $\sigma_{2}(\om)$, which is another
source of the output error bars. If we assume that it is just
random noise with the standard deviations at each frequency
given by $\delta\sigma_{1,j}$ and $\delta\sigma_{2,j}$ then the
standard deviation of the output of Eq.(\ref{Wlinsum}) is:
\begin{eqnarray}
F^{2}=\sum_{\nu=1}^{2}\sum_{j=1}^{N}u_{\nu,j}^2(\delta\sigma_{\nu,j})^2.
\end{eqnarray}
Our experience shows that the optimization of $D_{int}^{2}$
alone may provide very inaccurate results for a noisy input.
Instead, the minimization of a compound functional $D_{int}^2+w
F^{2}$, where $w$ is a weighting coefficient which is discussed
later, works much better. From a numerical point of view,
adding $w F^{2}$ enhances the diagonal elements of the matrix
$A$:
\begin{eqnarray}
A_{\nu\nu,ii}\rightarrow
A_{\nu\nu,ii}+w(\delta\sigma_{\nu,i})^2\nonumber
\end{eqnarray}

\noindent and makes it better conditioned. This addition
prevents the u-coefficients from growing too much, therefore
one can consider it as a {\em regularization} term.

The coefficient $w$ describes the relative significance of the
data noise compared to the inaccuracy of the linear
Eq.(\ref{Wlinsum}). The subtlety is that the optimal value of
$w$ is determined by $W(\omc)$ and is thus not known {\em a
priori}. A way out is to use a second optimization loop for
$w$. As a criterion, it is logical to minimize the estimated
total error bar $(\delta W)^{2}$, {\em i.e.} caused by both the
formula uncertainty and the input noise.

A simple estimate that we use (which can be perhaps improved)
is the following. For a given set of the u-coefficients, one
can find $W_{a}$, $F^{2}$ and $D_{int}^{2}$. Since the two
error sources are independent, one can approximately determine
the range of possible values of the true spectral weight $W$ by
the inequality:
\begin{equation}\label{RangeW}
(W-W_{a})^{2}<F^{2}+\kappa D_{int}^{2}W^{2}/\omc,
\end{equation}

\noindent where $\kappa$ is a number of the order 1-10. The
goal of this parameter is to adjust the accuracy of the rough
estimate (\ref{RangeW}). The variation of $\kappa$ by one order
of magnitude does not significantly affect the value of
$W_{0}$, although it does modify $\delta W$. We found that
$\kappa=5$ gives the best results in the numerical tests
described in the main text. It is likely that more accurate
estimates can remove an ambiguity here.

The inequality (\ref{RangeW}) can be resolved with respect to
$W$
\begin{equation}
W_{0}-\delta W < W < W_{0}+\delta W.
\end{equation}

\noindent where ($b=\kappa D_{int}^{2}/\omc$):
\begin{eqnarray}
W_{0}&=&\frac{W_{a}}{1-b}\nonumber \\
\delta W &=&\frac{\sqrt{b W_{a}^{2}+(1-b)F^{2}}}{1-b}.\nonumber
\end{eqnarray}

\noindent After the numerical minimization of $(\delta W)^{2}$
as a function of $w$, the u-coefficients become dependent on
$\sigma_{1,j}$ and $\sigma_{2,j}$. For example, in a case of a
very large value of $W$, the relevant importance of the input
error bars is small and the optimal $u$-coefficients tend to
oscillate stronger than in a case of small $W$.

\begin{figure}[htb]
   \centerline{\includegraphics[width=9cm,clip=true]{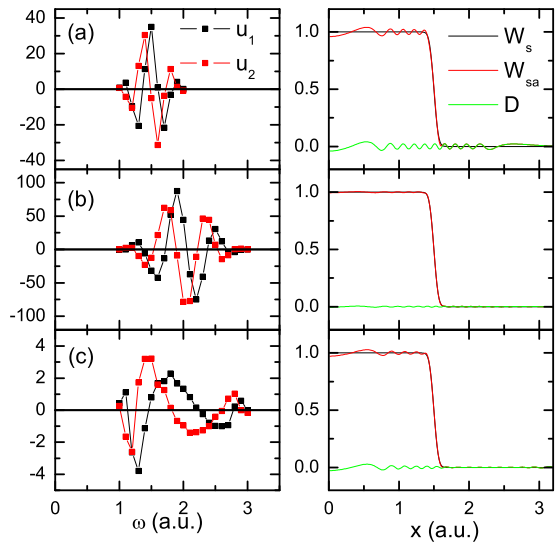}}
   \caption{The optimized u-coefficients (left panel) and the corresponding functions $\tilde{W}_{s}(x)$,
   $\tilde{W}_{s,a}(x)$ and $D(x)=\tilde{W}_{s}(x) - \tilde{W}_{s,a}(x)$ (right panel). Everywhere $\omc$=1.5,
   $\omin$=1,
   $\delta(\om)$=0.1. (a) $\omax$=2, no data noise; (b) $\omax$=3, no data noise (c) $\omax$=3, some data noise is
   added.}
   \label{Fig4}
\end{figure}

Fig. \ref{Fig4} shows some examples of the u-coefficients
optimized as described above and the corresponding discrepancy
function $D(x)$ for different sets of experimental frequencies
and noise levels. Clearly, the spectral discrepancy function
can be made quite small for all $x$ by a proper optimization.
The comparison between panels (a) and (b) tells that the
discrepancy function is smaller if the range of experimental
frequencies is broader. On the other hand, from panels (b) and
(c) one concludes that taking error bars into account makes the
u-coefficients much smaller, which results in a somewhat larger
$D(x)$ but in a better overall accuracy $(\delta W)^{2}$ (not
shown).

\

\section{Appendix 2. The equivalence of the least-square fit of
$\sigma_{1}(\om)$ and $\sigma_{2}(\om)$ and the linear
approximation formula}\label{Section2}

\

The most straightforward way to determine spectral weight is to
fit experimental data with a model function
\begin{eqnarray}\label{model}
\sigma_{mod}(\om)=\sum_{k=1}^{M}c_{k}\tilde{\sigma}_{k}(\om).
\end{eqnarray}
\noindent This is a linear superposition of some basis
functions
$\tilde{\sigma}_{k}(\om)=\tilde{\sigma}_{1,k}(\om)+i\tilde{\sigma}_{2,k}(\om)$,
Lorentzians, for example, each of which satisfies the KK
relations. Then one can estimate the partial sum-rule integral,
using this model:
\begin{eqnarray}\label{Wappoxc}
W(\omc)\approx
\int_{0}^{\omc}\sigma_{1,mod}(\om)d\om=\sum_{k=1}^{M}c_{k}\tilde{W}_{k},
\end{eqnarray}

\noindent where
$\tilde{W}_{k}=\int_{0}^{\omc}\tilde{\sigma}_{1,k}(\om)d\om$.

Fitting the data points
$\{\om_{j},\sigma_{\nu,j},\delta\sigma_{\nu,j}\}$ ($\nu=1,2$,
$j=1..N$) with the model function (\ref{model}) in the
least-square sense means the minimization of

\begin{equation}
\chi^{2}=\sum_{\nu=1}^{2}\sum_{j=1}^{N}\frac{1}{(\delta\sigma_{\nu,j})^{2}}\left[\sigma_{\nu,j}-
\sum_{k=1}^{M}c_{k}\tilde{\sigma}_{\nu,k}(\om_{j})\right]^{2}\nonumber
\end{equation}

\noindent with respect to the coefficients $c_{k}$. This gives
a linear system of equations

\begin{equation}\label{solutionc}
\sum_{k=1}^{M}a_{lk}c_{k}=b_{l}, \mbox{ or }
c_{k}=\sum_{l=1}^{M}(a^{-1})_{lk}b_{l},
\end{equation}

\noindent where

\noindent
\begin{eqnarray}\label{ab}
a_{lk}&=&\sum_{\nu=1}^{2}\sum_{j=1}^{N}
\frac{\tilde{\sigma}_{\nu,l}(\om_{j})\tilde{\sigma}_{\nu,k}(\om_{j})}{(\delta\sigma_{\nu,j})^{2}}\nonumber\\
b_{l}&=&\sum_{\nu=1}^{2}\sum_{j=1}^{N}
\frac{\tilde{\sigma}_{\nu,l}(\om_{j})\sigma_{\nu,j}}{(\delta\sigma_{\nu,j})^{2}}.
\end{eqnarray}

Substituting Eq.(\ref{solutionc}) and (\ref{ab}) into Eq.
(\ref{Wappoxc}) we obtain, after some transformations, an
expression identical to Eq.(\ref{Wlinsum}):

\begin{equation}
W(\omc)\approx
\sum_{\nu=1}^{2}\sum_{j=1}^{N}u_{\nu,j}\sigma_{\nu,j}
\end{equation}
\noindent with the u-coefficients independent of
$\sigma_{\nu,j}$:
\begin{eqnarray}\label{ufit}
u_{\nu,j}=\frac{1}{(\delta\sigma_{\nu,j})^{2}}\sum_{l,k=1}^{M}(a^{-1})_{l,k}\tilde{\sigma}_{\nu,k}(\om_{j})\tilde{W}_{l}.
\end{eqnarray}

This implies that the strategy to obtain $W(\omc)$ by fitting
$\sigma_{1}(\om)$ and $\sigma_{2}(\om)$ is nothing else but the
application of the linear formula (\ref{Wlinsum}) with specific
u-coefficients determined by the basis of functions
$\tilde{\sigma}_{k}(\om)$.


\begin{references}
\bibitem{RemarkCarleman}
The exact conditions on $\sigma(\omega)$ for the validity of the Carleman-Goluzin-Krylov formula are given in Ref.\cite{AizenbergKluwer93}.
For the present discussion it is important that all physical functions with the poles below the real axis would fit these requirements. In the case of
the optical conductivity of a superconductor, which has a pole exactly at the real axis, one can shift the pole slightly down. This
would only broaden the $\delta$-peak by an arbitrary small amount but simultaneously provide the necessary condition.
\bibitem{units} In these tests, $\om$ and $\sigma$
are taken in their respective (arbitrary) units; the unit of
$W(\omc)$ is the product of the two.
\end{references}
\end{document}